\documentclass[
preprint,
superscriptaddress,
groupedaddress,
amsmath,amssymb,
aps,
longbibliography,
pre
]{revtex4-2}

\usepackage{graphicx}
\usepackage{dcolumn}
\usepackage{xcolor}
\usepackage{bm}
\usepackage[title]{appendix}
\usepackage{lipsum}
\usepackage{float}

\usepackage{hyperref}
\hypersetup{
    colorlinks=true,
    linkcolor=blue,
    filecolor=magenta,      
    urlcolor=cyan,
    citecolor=red,
}
\begin{document}

\title{Structural Requirements for Ion-Acoustic Double Layers: A Parametric Perturbation Analysis of the Maxwellian Limit}

\author{Hamdi M. Abdelhamid}
\email{hamdi_mtprg@mans.edu.eg}
\author{Abeer A. Mahmoud}
 \affiliation{ Physics Department, Faculty of Science, Mansoura University, Mansoura, 35516 Egypt.}
\author{Naoki Sato}
\affiliation{National Institute for Fusion Science,
322-6 Oroshi-cho, Toki, Gifu 509-5292, Japan}
\affiliation{Graduate School of Frontier Sciences, The University of Tokyo,
Kashiwa, Chiba 277-8561, Japan}
\begin{abstract}

Standard Maxwellian plasmas exhibit a mathematical \textit{rigidity}, possessing insufficient degrees of freedom to support electrostatic double layers (DLs) and yielding only soliton solutions. This study investigates the hypothesis that the formation of DLs is a generic consequence of breaking this structural rigidity through parametric perturbation. By introducing two independent continuous control parameters, $\delta_1$ and $\delta_2$, into the electron distribution, we demonstrate that DLs are a structural property of any plasma model that relaxes the strict Maxwellian constraint. Through a Gardner small-amplitude expansion, we analytically prove that a perturbation must modify both the quadratic and cubic density coefficients to decouple the nonlinear structure and generate physical, supersonic double layers, deriving small-amplitude acoustic-limit threshold conditions of $\delta_1 > 1$ and $\delta_2 > 7/3$. We show that these theoretical boundaries broaden for large-amplitude, nonlinear structures. By mapping the exact existence regions of DLs in phase space, we demonstrate how higher-order terms relax the weak-amplitude limits, confirming that the Maxwellian state represents a singular point where the DL solution collapses.
\end{abstract}

\maketitle

\section{Introduction}

Electrostatic double layers (DLs)---monotonic potential structures connecting distinct plasma states---are fundamental to particle acceleration and energy dissipation in both auroral and laboratory plasmas \cite{Block1977, Reddy1991, Hershkowitz1985, Chan1984, Charles2007, Saha2023, Alex2017}. The macroscopic dynamics of these structures have been further highlighted by recent laboratory experiments, which demonstrate the spontaneous emergence of current-free DLs in expanding plasmas \cite{Charles2007, Saha2023}, as well as complex sequences of multiple double layers (MADLs) triggered by current-driven instabilities in glow discharge and helicon radio-frequency (RF) setups \cite{Alex2017, Saha2023}. However, describing DLs using standard fluid models remains challenging. It is well established that a simple plasma consisting of cold ions and Boltzmann-distributed (Maxwellian) electrons possesses insufficient degrees of freedom to satisfy the boundary conditions required for a DL, admitting only compressive soliton solutions \cite{Bharuthram1986}.

To resolve this discrepancy, theoretical investigations have typically invoked specific microscopic kinetic effects, such as trapped particle phase-space vortices \cite{Schamel1972, Schamel1986}, or specific non-thermal particle distributions like Kappa or Tsallis statistics \cite{Hellberg2002, Hatami2025}. Recent models often utilize multiple non-thermal parameters (e.g., plasmas with two distinct nonextensive electron species) to accurately map the intricate boundaries of DL existence \cite{Hatami2025}. Furthermore, analytical treatments have demonstrated that standard weakly nonlinear expansions (e.g., Korteweg-de Vries equations) fail to capture the solitary wave-to-double layer transition \cite{Rosenau1988}. While these specific kinetic and multi-parameter non-thermal models yield DLs, their reliance on model-specific nonlinearities can obscure the fundamental mathematical mechanism of DL formation.

In this paper, it is demonstrated that the emergence of DLs is not strictly dependent on specific families of non-thermal distributions, but is a generic consequence of deviating from the Maxwellian limit. By introducing independent parametric perturbations ($\delta_1$ and $\delta_2$) to the lowest-order polynomial expansion of the electron density, these disparate physical models can be unified under a single macroscopic framework. We establish that the Maxwellian distribution represents a degenerate structural state, and that decoupling the quadratic and cubic density nonlinearities is a structural requirement for the existence of stationary, supersonic DLs across both weak and large amplitude regimes.

The paper is organized as follows. In Section \ref{Classical}, we review the fluid framework and Sagdeev pseudopotential formalism to demonstrate the mathematical limitation of the Maxwellian distribution. Section \ref{PP} introduces the dual-parameter perturbation model, establishes the requirement for independent quadratic and cubic density corrections, and derives the small-amplitude expansion coefficients. In Section \ref{SS}, we analyze the structural stability and derive the small-amplitude acoustic-limit threshold conditions required for double layer formation. Section \ref{Numerical} evaluates the exact numerical limits and maps the existence domain for large-amplitude dynamics. Finally, Section \ref{Conclusion} provides a discussion of the theoretical implications and concludes the study.

\section{Classical Analysis of Distribution Functions} \label{Classical}

Before introducing the parametric model, we briefly revisit the classical Sagdeev formalism to demonstrate why the Maxwellian distribution fails to support double layers and how generalized distributions resolve this.

\subsection{Governing Dynamical Equations}
We consider a one-dimensional, collisionless, unmagnetized plasma consisting of cold fluid ions and an arbitrary population of electrons. The dynamics of the system are governed by the ion continuity and momentum equations, coupled with Poisson's equation for the electrostatic potential \cite{Sagdeev1966, Chen2015}:

\begin{align}
    \frac{\partial n_i}{\partial t} + \frac{\partial (n_i u_i)}{\partial x} &= 0 \label{eq:continuity} \\
    \frac{\partial u_i}{\partial t} + u_i \frac{\partial u_i}{\partial x} &= -\frac{\partial \phi}{\partial x} \label{eq:momentum} \\
    \frac{\partial^2 \phi}{\partial x^2} &= n_e(\phi) - n_i \label{eq:poisson}
\end{align}

Here, $n_i$ and $n_e$ are the ion and electron number densities normalized by the equilibrium density $n_0$. The ion fluid velocity $u_i$ is normalized by the ion sound speed $C_s = (k_B T_e/m_i)^{1/2}$, the electrostatic potential $\phi$ is normalized by $k_B T_e/e$, time $t$ by the inverse ion plasma frequency $\omega_{pi}^{-1}$, and space $x$ by the Debye length $\lambda_D$. We assume charge neutrality at equilibrium, such that $n_{e0} = n_{i0} = 1$ and $u_{i0}= 0$ as $|x| \to \infty$. 

\subsection{General Sagdeev Formalism}
To find stationary wave solutions, the system is transformed to a moving frame $\xi = x - Mt$, where $M$ is the Mach number (wave speed normalized by $C_s$). 
Applying the boundary conditions $n_i \to 1$, $u_i \to 0$, and $\phi \to 0$ as $\xi \to \pm \infty$, the continuity and momentum equations yield the ion density:
\begin{equation}
    n_i(\phi) = \frac{1}{\sqrt{1 - \frac{2\phi}{M^2}}}
\end{equation}
Substituting this into Poisson's equation (\ref{eq:poisson}) and integrating once with respect to $\phi$ yields the energy integral:
\begin{equation}
    \frac{1}{2}\left(\frac{d\phi}{d\xi}\right)^2 + V(\phi) = 0
\end{equation}
where the Sagdeev potential $V(\phi)$ is defined by integrating the charge density \cite{Sagdeev1966}. Noting that the equilibrium conditions dictate $V(0)=0$, the general form is:
\begin{equation}
    V(\phi) = M^2 \left( 1 - \sqrt{1 - \frac{2\phi}{M^2}} \right) - \int_0^\phi n_e(\psi) d\psi
    \label{eq:general_sagdeev}
\end{equation}
This generalized approach relies on retaining the full Poisson equation, as assuming strict quasi-neutrality can yield different structural results for finite-amplitude double layers \cite{Roychoudhury1989}. For a double layer of finite amplitude $\phi_m \neq 0$ to exist, two simultaneous boundary conditions must be met \cite{verheest2012waves, Bandyopadhyay2000}:
\begin{align}
    V(\phi_m) &= 0 \quad \text{(Energy Conservation)} \label{eq:cond1} \\
    V'(\phi_m) &= 0 \quad \text{(Charge Neutrality)} \label{eq:cond2}
\end{align}

\subsection{Non-existence in Maxwellian Plasmas}
For Maxwellian electrons, $n_e(\phi) = e^\phi$. Substituting this into conditions (\ref{eq:cond1}) and (\ref{eq:cond2}) yields a system of algebraic equations. From $V'(\phi_m) = n_i(\phi_m) - n_e(\phi_m) = 0$, we find the relation $M^2 = 2\phi_m / (1 - e^{-2\phi_m})$. Substituting this back into the energy condition provides the transcendental equation:
\begin{equation}
    1 - 2\phi_m e^{-\phi_m} - e^{-2\phi_m} = 0
    \label{eq:maxwellian_fail}
\end{equation}
Analytical inspection of Eq. (\ref{eq:maxwellian_fail}) indicates that the only real solution is the trivial root $\phi_m = 0$. Thus, a plasma with cold ions and strictly Boltzmann electrons is unable to sustain a double layer structure \cite{Bharuthram1986}; the single parameter $M$ cannot simultaneously satisfy both boundary conditions for any non-zero amplitude.

\section{The Parametric Perturbation Model} \label{PP}

It is proposed that DL emergence depends on the decoupling of nonlinear polynomial terms from the linear dispersion relation. To motivate the choice of adding specific quadratic and cubic corrections, the model can be regarded as a low-order Taylor perturbation of the Maxwellian electron response. Namely, one may start from the unperturbed state,
\begin{equation}
    n_{e}^{(0)}(\phi) = e^{\phi} = \sum_{j=0}^{\infty}\frac{\phi^j}{j!},
\end{equation}
and introduce a generic perturbation $\Delta n_e(\phi) = \sum_{j=0}^{\infty}a_j\phi^j$. The total electron density then becomes
\begin{align}
    n_e(\phi) &= n_e^{(0)}(\phi) + \Delta n_e(\phi) \nonumber \\
    &= 1 + a_0 + (1 + a_1)\phi + \left(\frac{1}{2} + a_2\right)\phi^2 + \left(\frac{1}{6} + a_3\right)\phi^3 + \mathcal{O}(\phi^4).
\end{align}
The constant correction $a_0$ changes the equilibrium density and must be set to zero to preserve the boundary condition $n_e(0)=1$. The linear correction $a_1$ only modifies the linear response---equivalently renormalizing the acoustic speed and Debye length scaling---and does not affect the nonlinear mechanisms discussed here. Therefore, after fixing the equilibrium and linear Maxwellian responses, the first genuinely nonlinear corrections occur at the quadratic and cubic orders. In this sense, the structural deviation from thermodynamic equilibrium can be interpreted through the minimal two-parameter truncation:
\begin{equation}
    n_e(\phi) = e^\phi + \delta_1\phi^2 + \delta_2\phi^3,
    \label{eq:density}
\end{equation}
where $\delta_1 = a_2$ and $\delta_2 = a_3$ independently perturb the quadratic and cubic nonlinear coefficients. Setting $\delta_1 = \delta_2 = 0$ recovers the classical equilibrium state. Furthermore, because this truncated expansion acts as a phenomenological model, any physically valid solution must explicitly satisfy the macroscopic admissibility condition $n_e(\phi) > 0$ across the entire domain of the localized potential structure. For the parameter regimes and finite amplitudes evaluated in this study, this positivity constraint is rigorously maintained.

\subsection{The Requirement for Independent Quadratic and Cubic Terms}
The form in Eq. (\ref{eq:density}) employs two independent control parameters, which provides sufficient degrees of freedom to prevent the double-root boundary conditions from becoming overconstrained.

\textbf{1. The Physical Requirement (The Quadratic Term $\delta_1$):}
The quadratic perturbation ($\delta_1 \phi^2$) relates to non-thermal plasmas. For instance, expanding a Kappa distribution \cite{Hellberg2002, Baluku2008} yields:
\begin{equation}
    n_\kappa(\phi) \approx 1 + \phi + \left( \frac{1}{2} + \frac{1}{2\kappa - 3} \right)\phi^2 + \mathcal{O}(\phi^3)
\end{equation}
Comparing the $\mathcal{O}(\phi^2)$ terms to Eq. (\ref{eq:density}), we identify $\delta_1 \sim \frac{1}{2\kappa - 3}$, which decouples the linear shielding length from the primary nonlinear steepening. This is consistent with earlier generalized equations of state \cite{Goswami1985} and kinetic theories of phase space vortices, where two-parametric shifts have recently been shown to be fundamental for resolving mathematically undisclosed solitary structures \cite{Schamel1982, Schamel1983, Schamel2020}. By utilizing an independent parameter $\delta_1$, the model avoids linking the primary nonlinearity to higher-order structural constraints, acting as a proxy for arbitrary non-equilibrium distributions \cite{Kim1983}.

\textbf{2. The Mathematical Requirement (The Cubic Term $\delta_2$):}
However, a purely quadratic perturbation ($\delta_2 = 0$) confines the Mach number to the sub-acoustic regime ($M < 1$). As shown by Rosenau \cite{Rosenau1988}, capturing the critical threshold where solitary waves transition into double layers requires extending beyond weakly nonlinear scaling constraints. The independent cubic term ($\delta_2 \phi^3$) modifies the sub-dominant quartic potential coefficient, allowing solutions to act as physical potential wells ($V < 0$) traveling at supersonic speeds ($M > 1$).

\subsection{The Sagdeev Potential}
For the dual-perturbed density defined in Eq. (\ref{eq:density}), integrating the general Sagdeev potential yields:
\begin{equation}
    V(\phi) = M^2 \left( 1 - \sqrt{1 - \frac{2\phi}{M^2}} \right) - \left( e^\phi - 1 + \delta_1 \frac{\phi^3}{3} + \delta_2 \frac{\phi^4}{4} \right)
\end{equation}

\subsection{Small Amplitude Expansion}
To evaluate the onset conditions for DL formation, we examine the small-amplitude limit ($|\phi| \ll 1$). In this weakly nonlinear regime, extending the classical Korteweg-de Vries analysis to include cubic nonlinearities yields the Gardner equation (the combined KdV-modified KdV equation) \cite{Gardner1967, Miura1968}. Expanding the complete pseudopotential $V(\phi)$ into a Taylor series up to the fourth order provides the corresponding stationary-wave pseudo-energy integral,
\begin{equation}
    V(\phi) \approx A \phi^2 + B \phi^3 + C \phi^4
\end{equation}
Subtracting the series expansions of the electron density terms from the expanded ion fluid terms provides the structural coefficients as functions of the Mach number $M$ and the control parameters $\delta_1, \delta_2$ \cite{AliShan2014, Sahu2011}:
\begin{align}
    A(M) &= \frac{1}{2} \left( \frac{1}{M^2} - 1 \right) \label{eq:coeffA} \\
    B(M, \delta_1) &= \frac{1}{2M^4} - \frac{1}{6} - \frac{\delta_1}{3} \label{eq:coeffB} \\
    C(M, \delta_2) &= \frac{5}{8M^6} - \frac{1}{24} - \frac{\delta_2}{4} \label{eq:coeffC}
\end{align}

\section{Parametric Existence Conditions}\label{SS}

The existence of a rarefactive DL ($\phi_m < 0$) depends on three constraints within the Sagdeev potential roots.

\begin{enumerate}
    \item \textbf{Supersonic Condition:} To avoid strong Landau damping and ensure separation from the linear wave spectrum, any physical solitary wave must propagate faster than the relevant acoustic speed ($M > 1$) \cite{Chen2015, Washimi1966, Witt1986}. From Eq. (\ref{eq:coeffA}), $M > 1$ dictates that the dispersive coefficient is negative ($A < 0$) \cite{Tagare2000}.
    \item \textbf{Double Root Geometry:} A double root at finite amplitude $\phi_m$ requires the discriminant of the Sagdeev potential expansion to vanish. This imposes the geometric constraint $B^2 - 4AC = 0$, which yields $4AC = B^2$. Since $B^2 \ge 0$, the structure requires the product $4AC > 0$. Since the physical propagation condition limits $A < 0$, the existence of a double root restricts the quartic coefficient to negative values ($C < 0$) \cite{verheest2012waves, Sahu2011, Bandyopadhyay2000}.
    \item \textbf{Rarefactive Constraint:} The root resides at $\phi_m = -B / (2C)$. For a rarefactive well ($\phi_m < 0$) with $C < 0$, the primary nonlinear steepening coefficient must be negative ($B < 0$) \cite{Chan1984}.
\end{enumerate}

Our physical analysis focuses exclusively on rarefactive structures ($\phi_m < 0$). For compressive structures ($\phi > 0$), the cold ion density $n_i(\phi) = (1 - 2\phi/M^2)^{-1/2}$ encounters a fundamental singularity at $\phi = M^2/2$, corresponding to ion wave breaking and total particle reflection \cite{Bharuthram1986}. Thus, strictly stationary compressive fluid structures cannot be sustained in this baseline cold-ion framework without introducing additional physics, such as finite ion temperature, secondary ion beams \cite{Bharuthram1986, Chatterjee1994, Paul2020}, negative ions \cite{Mishra2002}, or positron populations \cite{Mishra2007, Saini2013}.

\subsection{The Failure of Purely Quadratic Perturbations}
Consider a model where the density perturbation is purely quadratic: $n_e = e^\phi + \delta_1 \phi^2$ (i.e., $\delta_2 = 0$). In this scenario, the quartic potential coefficient $C$ remains independent of the structural perturbation. Evaluating Eq. (\ref{eq:coeffC}) with $\delta_2 = 0$ near the acoustic limit ($M \approx 1$) yields:
\begin{equation}
    C \approx \frac{5}{8} - \frac{1}{24} = \frac{14}{24} = \frac{7}{12}
\end{equation}
Because this model locks $C$ to a strictly positive value ($C > 0$), the geometric double-root condition ($4AC > 0$) forces $A > 0$. However, $A > 0$ implies $1/M^2 - 1 > 0$, structurally confining any double-root solutions to the sub-acoustic regime ($M < 1$). Physically, such sub-acoustic structures yield positive potential domains ($V > 0$), acting as unphysical reflective barriers rather than propagating transmissive wells.

\subsection{Small-Amplitude Acoustic-Limit Threshold Conditions}
By extending the perturbation to include an independent cubic density term ($\delta_2 \phi^3$), the quartic potential coefficient $C$ decouples from its rigid, positive Maxwellian value. As derived in Eq. (\ref{eq:coeffC}), near the acoustic speed ($M \to 1$):
\begin{equation}
    C \approx \frac{5}{8} - \frac{1}{24} - \frac{\delta_2}{4} = \frac{7}{12} - \frac{\delta_2}{4}
\end{equation}
To satisfy the condition required for supersonic structures ($C < 0$), the secondary perturbation parameter must exceed:
\begin{equation}
    \frac{\delta_2}{4} > \frac{7}{12} \implies \delta_2 > \frac{7}{3} \approx 2.33
\end{equation}
Simultaneously, evaluating the primary nonlinearity for a rarefactive well ($B < 0$) near $M \approx 1$ gives $B \approx 1/2 - 1/6 - \delta_1/3 = \frac{1}{3}(1 - \delta_1)$, which requires $\delta_1 > 1$.

Because the structural parameters decouple the equations, they must satisfy their respective conditions independently. Thus, forming a double layer near the acoustic limit ($M \to 1, \phi_m \to 0$) requires independent quadratic and cubic perturbations exceeding the distinct thresholds $\delta_1 > 1$ and $\delta_2 > 7/3$. These bounds are best described as small-amplitude or acoustic-limit threshold conditions, rather than global existence thresholds, since the large-amplitude analysis shows that the bounds are relaxed, see section \ref{Numerical}. This $7/3$ value acts as the upper asymptotic bound of the required threshold spectrum, analogous to critical non-thermal concentration limits observed in multi-component plasmas \cite{Hatami2025, AliShan2021}.

\section{Exact Numerical Limits and Large-Amplitude Dynamics}\label{Numerical}

While the Gardner expansion identifies the polynomial decoupling necessary for DL formation ($\delta_1 > 1, \delta_2 > 7/3$), this truncation is strictly valid only in the weakly nonlinear limit ($|\phi_m| \ll 1, M \to 1$). To accurately map the existence domain for large-amplitude structures, the exact mathematical conditions must be evaluated using the un-truncated Sagdeev potential to determine the permissible Mach number \cite{ElTantawy2011, Saini2013}. Evaluating the full potential is crucial, as it can reveal complex structural features and a broader parameter space for existence that weak-amplitude expansions miss \cite{Varghese2022}.

The lower bound for solitary structures remains the acoustic speed ($M_{min} = 1$). A solitary wave transitions into a DL at a specific Mach number $M$, which occurs at a maximum amplitude $\phi_m$ satisfying the double-root conditions:
\begin{align}
    V(\phi_m) &= 0, \label{eq:dl_1} \\
    \left. \frac{dV}{d\phi} \right|_{\phi=\phi_m} &= 0, \label{eq:dl_2} \\
    \left. \frac{d^2V}{d\phi^2} \right|_{\phi=\phi_m} &< 0. \label{eq:dl_3}
\end{align}

Enforcing charge neutrality (Eq.~\ref{eq:dl_2}) at $\phi_m$ provides the Mach number explicitly as a function of the maximum amplitude and structural parameters:
\begin{equation}
    M = \sqrt{ \frac{2\phi_m}{1 - \left( e^{\phi_m} + \delta_1 \phi_m^2 + \delta_2 \phi_m^3 \right)^{-2}} }.
    \label{eq:exact_M}
\end{equation}
Substituting Eq.~(\ref{eq:exact_M}) into the energy conservation condition (Eq.~\ref{eq:dl_1}) yields the exact transcendental equation defining the geometric bifurcation boundary:
\begin{equation}
    \frac{2\phi_m \left( e^{\phi_m} + \delta_1 \phi_m^2 + \delta_2 \phi_m^3 \right)}{e^{\phi_m} + \delta_1 \phi_m^2 + \delta_2 \phi_m^3 + 1} - \left( e^{\phi_m} - 1 + \frac{\delta_1 \phi_m^3}{3} + \frac{\delta_2 \phi_m^4}{4} \right) = 0.
    \label{eq:exact_DL_condition}
\end{equation}

For a fixed primary perturbation parameter ($\delta_1$), Eq.~\eqref{eq:exact_DL_condition} is solved numerically to determine the critical cubic perturbation ($\delta_2$) required for a prescribed maximum amplitude ($\phi_m$). The corresponding Mach number $M$ is subsequently obtained from Eq.~\eqref{eq:exact_M}. This procedure rigorously maps the exact geometric bifurcation boundary without relying on weak-amplitude approximations.

The data in Table~\ref{critical_values} demonstrates how larger amplitudes relax the external perturbation requirement. For a relatively shallow rarefactive well ($\phi_m \approx -0.8$), the exact boundary requires $\delta_2 \approx 1.35$, falling substantially below the asymptotic weak-amplitude bound of $\delta_2 > 7/3 \approx 2.33$. As the well deepens to $\phi_m \approx -1.2$, the required $\delta_2$ decreases monotonically to $\approx 0.96$---a reduction of nearly 60\% from the acoustic limit. This divergence confirms that the strict $\delta_2 > 7/3$ threshold is a \emph{necessary} condition only in the weakly nonlinear acoustic limit ($|\phi_m| \to 0$); it is never a sufficient condition, nor is it a necessary global constraint for large-amplitude structures. The noticeable broadening of these wells at larger amplitudes clearly illustrates how the exact, higher-order nonlinearities naturally step in to compensate for the reduced $\delta_2$ requirement.
\begin{table}[htbp]
\caption{Exact numerical boundaries for the formation of large-amplitude rarefactive double layers at varying maximum amplitudes, given a fixed primary perturbation $\delta_1 = 1.5$.}
\label{critical_values}
\begin{ruledtabular}
\begin{tabular}{ccc}
Maximum Amplitude ($\phi_m$) & Mach Number ($M$) & Cubic Perturbation ($\delta_2$) \\
\colrule
$-0.80$ & $1.30176$ & $1.35185$ \\
$-1.00$ & $1.55113$ & $1.12891$ \\
$-1.20$ & $2.03024$ & $0.964239$ \\
\end{tabular}
\end{ruledtabular}
\end{table}

\begin{figure}[H]
  \centering
  \includegraphics[width=0.58\textwidth]{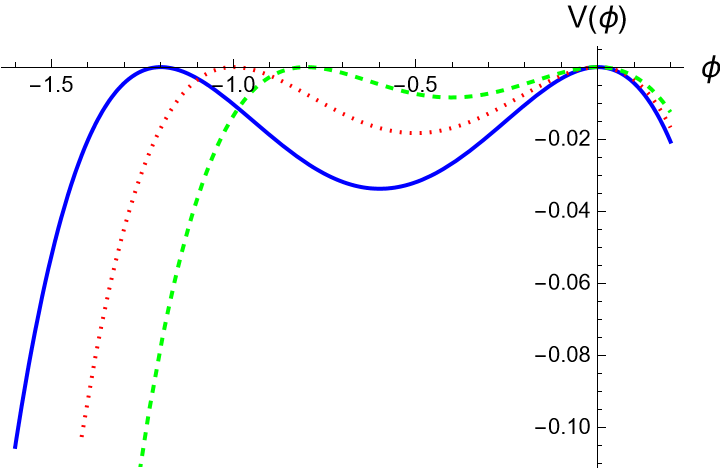}
  \caption{The exact un-truncated Sagdeev pseudopotential $V(\phi)$ versus the electrostatic potential $\phi$ for large-amplitude double layers at a fixed primary perturbation $\delta_1 = 1.5$. The curves illustrate the formation of exact double roots satisfying $V(\phi_m)=0$ and $V'(\phi_m)=0$ for maximum amplitudes $\phi_m \approx -1.2$ (solid blue line, $M \approx 2.030, \delta_2 \approx 0.964$), $\phi_m \approx -1$ (dotted red line, $M \approx 1.551, \delta_2 \approx 1.129$), and $\phi_m \approx -0.8$ (dashed green line, $M \approx 1.302, \delta_2 \approx 1.352$).}
  \label{fig:sagdeev_dl}
\end{figure}

Figure~\ref{fig:sagdeev_dl} displays the un-truncated Sagdeev potentials corresponding to the exact solutions from Table~\ref{critical_values}. These curves exhibit tangential double roots at $\phi_m = -1.2, -1.0,$ and $-0.8$, identically satisfying both $V(\phi_m)=0$ and $V'(\phi_m)=0$. Furthermore, the negative curvature at these roots ($V''(\phi_m)<0$, evident from the concave-downward structure) ensures they represent physically admissible, transmissive potential wells. The noticeable broadening of these wells at larger amplitudes clearly illustrates how the exact, higher-order nonlinearities naturally step in to compensate for the reduced $\delta_2$ requirement.

Crucially, as the amplitude deepens, the required cubic perturbation parameter ($\delta_2$) decreases monotonically. The exact numerical boundaries ($\delta_2 \approx 1.35, 1.13, 0.96$) fall significantly below the $\delta_2 > 7/3 \approx 2.33$ analytical threshold. Thus, the $7/3$ limit serves strictly as an asymptotic ceiling for vanishingly small amplitudes ($|\phi_m| \to 0$); large-amplitude structures rely less on higher-order perturbations to fulfill the geometric curvature constraints, occupying a significantly broader parameter space.
\begin{figure}[H]
  \centering
  \includegraphics[width=0.5\textwidth]{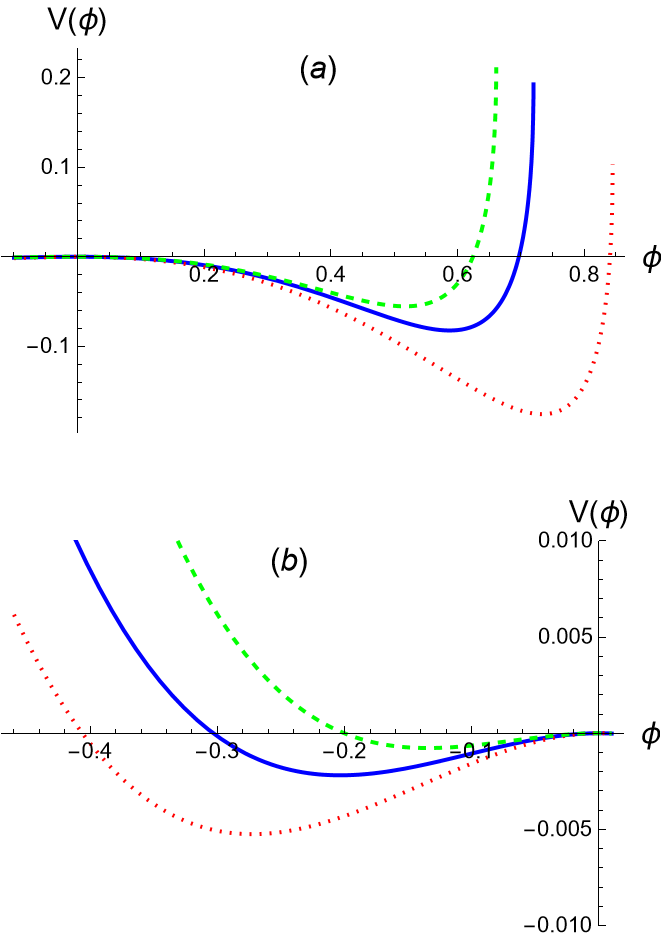}
   \caption{Exact Sagdeev pseudo-potential $V(\phi)$ demonstrating the coexistence of solitary waves. (a) Top panel: The potential profile on the positive $\phi$ axis, demonstrating robust compressive solitary roots. (b) Bottom panel: A magnified view on the negative $\phi$ axis, showing the simultaneous formation of shallow rarefactive potential wells for the exact same physical states. Evaluated parameter sets: $M=1.20, \delta_1=1.6, \delta_2=-0.1$ (blue solid); $M=1.30, \delta_1=1.4, \delta_2=-0.2$ (red dotted); and $M=1.15, \delta_1=2.0, \delta_2=-0.5$ (green dashed).}
  \label{fig:coexistence}
\end{figure}
Furthermore, the exact parametric model accommodates a versatile coexistence of solitary waves. While compressive DLs are excluded prior to the ion wave-breaking limit ($\phi = M^2/2$), Figure \ref{fig:coexistence} demonstrates that the un-truncated potential effortlessly supports both deep compressive solitons and shallow rarefactive solitons simultaneously. 

As shown in Figure \ref{fig:coexistence}, depending purely on initial excitation conditions, the exact same macroscopic plasma state can propagate either a robust compressive pulse or a small-amplitude rarefactive pulse. This structural asymmetry confirms that deviating from the Maxwellian state via decoupled quadratic and cubic perturbations comprehensively resolves the structural singularities of the equilibrium limit across all amplitude regimes.
Such coexistence is strictly forbidden in the standard Maxwellian limit, which is too mathematically rigid to produce physical, transmissive roots. The ability of the exact potential to support these diverse wave structures, even when $\delta_2 < 0$, reveals a fundamental principle: escaping the Maxwellian degeneracy simply requires the presence of independent quadratic and cubic degrees of freedom, rather than specific parameter magnitudes.

Finally, it must be noted that within this strictly cold-ion fluid framework, stationary compressive double layers remain prohibited up to the ion wave-breaking singularity ($\phi = M^2/2$), owing to the singular behavior of the ion density $n_i(\phi) = (1 - 2\phi/M^2)^{-1/2}$. This limitation, however, pertains exclusively to the compressive branch; the rarefactive DL branch remains universally accessible. Accessing the compressive DL branch necessitates introducing supplementary physics, such as warm ions, multi-ion species, or beam-plasma configurations \cite{Bharuthram1986, Chatterjee1994, Paul2020}. Nonetheless, the fundamental mechanism of geometric root formation via polynomial decoupling remains identically applicable.

\section{Discussion and Conclusion}\label{Conclusion}

In this paper, it is demonstrated that the inability of Maxwellian plasmas to sustain electrostatic double layers arises from a structural constraint in the governing equations. The strict Boltzmann relation enforces a coupling between the dispersive, nonlinear, and higher-order terms of the Sagdeev potential, creating an overconstrained system where the double-root boundary condition cannot be met at supersonic speeds \cite{Bharuthram1986}. 

By introducing generic parametric perturbations ($\delta_1$ and $\delta_2$) to both the quadratic and cubic orders of the electron density, these terms decouple, providing the degrees of freedom necessary for DL formation \cite{Rosenau1988, Schamel1986}. This dual-parameter approach provides a generalized framework that unifies disparate physical models. Whether the underlying physics involves superthermal acceleration (Kappa), long-range non-extensive correlation (Tsallis) \cite{Hatami2025}, or phase-space particle trapping \cite{Schamel1982, Schamel1983, Goswami2008}, these phenomena manifest as independent modifications to the polynomial coefficients of the density expansion. Consequently, different types of double layers appear as subclasses of the same structural condition.

Furthermore, modifying only the lowest-order nonlinearity (quadratic) restricts solutions to the sub-acoustic regime ($M < 1$). Extending the structural perturbation to an independent cubic order is necessary to decouple the quartic potential coefficient, allowing supersonic potential drops and the formation of exact double roots. The analytical small-amplitude acoustic-limit threshold conditions ($\delta_1 > 1, \delta_2 > 7/3$) determine the mathematical onset of DL formation. However, precise numerical integration of the un-truncated Sagdeev pseudopotential \cite{Mamun2010, Dalui2021} shows that these boundaries relax significantly as structures deepen into the nonlinear regime.

Ultimately, the unperturbed Maxwellian state remains a singular point where the DL solution collapses. The requirement for decoupled, higher-order deviations from thermodynamic equilibrium extends beyond standard electron-ion models. It governs the formation of electrostatic shocks in highly complex states, ranging from ultra-relativistic degenerate dusty plasmas \cite{Mamun2010} and planetary ionospheres \cite{Ahmed2020} to multi-ion species, expanding RF configurations, and beam-plasma laboratory environments \cite{Lakhina2014, Paul2020, Charles2007, Alex2017, Saha2023}. These higher-order structural deviations act as a universal geometric requirement for the existence of electrostatic double layers \cite{Block1977, Michelsen1982, Chan1984}.

\bibliography{DL_REF}

\end{document}